 \documentstyle[11pt]{article} 
\begin{document}
\begin{titlepage}
\title{Selective continuous quantum measurements:\\
       Restricted path integrals and wave equations
\thanks{\it submitted to Phys. Lett. A}}
\author{Lajos Di\'osi
\thanks{E-mail: diosi@rmki.kfki.hu}\\
KFKI Research Institute for Particle and Nuclear Physics\\
H-1525 Budapest 114, POB 49, Hungary\\\\
{\it bulletin board ref.: quant-ph/9501009}}
\date{}
\maketitle
\begin{abstract}
We discuss both the restricted path integral (RPI)
and the wave equation (WE) techniques in the theory of continuous quantum
measurements. We intend to make Mensky's fresh review complete by
transforming his "effective" WE with complex Hamiltonian into
Ito-differential equations.
\end{abstract}
\end{titlepage}
\baselineskip=23pt
\section{Introduction}

Quite recently, a short review on two mathematical techniques of
continuous quantum measurements has been presented by Mensky \cite{Men94}.
The basic ideas and formal elements of the restricted path integral (RPI)
method (Sect.2) and
the master equation (ME) method (Sect.3), respectively, have been explained.
Selective (Sect.4) and non-selective (Sect.5) measurements have been
discussed. The former, being the more complicated one, only has been given a
very brief outline anticipating an efficient method of
Schr\"odinger-equation replacing circuitous path integrals.
In the present Letter we bring out this efficient wave equation (WE)
which has emerged from a great body of earlier works.

The idea of RPI to model continuous measurements originates from 1979
\cite{Men79}. Elaborating of the {\it statistical} theory within RPI have been
promoted basically by Refs.~\cite{BarLanPro82CavMil87,Dio88Ito,Dio90},
in addition to the papers cited in Ref.~\cite{Men94}.
The proof of the correspondence between the RPI and
the Ito-stochastic WE formalism (5ab) was given first in 1988 \cite{Dio88Ito}.

\section{Selective measurements}

Let us start with recapitulating Mensky's RPI formalism \cite{Men94}.
In his notations,
the state vector of the system under continuous measurement evolves as
\hbox{$\psi_t^\alpha=U_t^\alpha\psi_0$}, where the propagator is expressed
by the following RPI:
\begin{equation}
U_t^\alpha(q^{\prime\prime},q^\prime)=
\int_\alpha d[p]d[q]\exp\left({i\over\hbar}\int_0^{~t}
[p\dot q-H(p,q,t)]dt\right).
\end{equation}
The output of continuous measurement is labelled by $\alpha$, and
its probability distribution is given by
\begin{equation}
P(\alpha)=\langle\psi_t^\alpha\vert\psi_t^\alpha\rangle.
\end{equation}
(Interestingly, neither this statement nor the fact that $\psi_t^\alpha$
is unnormalized do appear in Ref.~\cite{Men94} explicitly.)
For the simple case of continuous monitoring an observable $A(p,q,t)$,
the propagator's RPI reduces to unrestricted path integral:
\begin{eqnarray}
&{}&U_t^{[a]}(q^{\prime\prime},q^\prime)\nonumber\\
&=&\int d[p]d[q]\exp\left(
{i\over\hbar}\int_0^{~t}[p\dot q-H(p,q,t)]dt
	    -\kappa\int_0^{~t}[A(p,q,t)-a(t)]^2dt \right).
\end{eqnarray}
A comparison of this path integral with Feynman's standard ones leads to the
naive conjecture that
"one may forget about any path integrals and reduce the problem to
solving the Schr\"odinger-equation with a complex Hamiltonian" \cite{Men94}:
\begin{equation}
{\partial\over\partial t}\psi_t
=\left( -{i\over\hbar}H(p,q,t)-\kappa[A(p,q,t)-a(t)]^2 \right)\psi_t
\end{equation}
where a functional dependence of $\psi_t$ on
\hbox{$[a]\equiv\{a(t^\prime)\vert0\le t^\prime \le t\}$}
is understood though has not been denoted explicitly.

It is inevitable to note that
this Schr\"odinger-equation is unconventional:
it is {\it not} linear, norm-conserving, deterministic, and regular either.
Its stochasticity is obvious from the fact that the effective
Hamiltonian depends on the measurement
output record $[a]$ whose probability distribution still depends on the
state vector via the Eq.~(2): \hbox{$P[a]=\Vert\psi_t\Vert^2$}.
{\it Nevertheless, these coupled functional equations
can be disentangled into two separate stochastic equations: one for the
state vector and another for the measurement output \cite{Dio88Ito}.}
It turns out that both $\psi_t$ and $a(t)$ are Wiener-processes rather than
regular functions of $t$:
$$
{\partial\over\partial t}\psi_t=
\left( -{i\over\hbar}H(p,q,t)
       -{\kappa\over2}[A(p,q,t)-\langle A\rangle_t]^2
       +\sqrt{\kappa}[A(p,q,t)-\langle A\rangle_t]\dot\xi_t
\right)\psi_t
\eqno(5a)$$
$$
a(t)=\langle A\rangle_t+{1\over2\kappa}\dot\xi
\eqno(5b)$$
where $\langle A\rangle_t$ denotes the expectation value of the
observable $A(p,q,t)$ in the quantum state $\psi_t$. The "function" $\xi_t$
is the standard Wiener-process whose time-derivative $\dot\xi_t$ is the
standard white-noise with $\delta(t)$ as auto-correlation.

The Eqs.~(5ab) provide a radical improvement as compared to the naive
WE (4); explicit solutions become available for certain special
cases like, e.g., for continuous position measurement of free particles
\cite{Dio88Ito}.

\section{Concluding remarks}

It should be noted that, alternatively to RPI, a theory of quantum filtering
has been devised to model continuous measurement and resulted in equations
mathematically equivalent to (5ab) \cite{Bel89BarBel91BelSta92}.
Furthermore, investigating the so-called quantum measurement problem
have led to important components of the formalism
explained in Ref.~[1] and in the present Letter. Simple stochastic WE
of structure (5a) was found heuristically \cite{Gis84} without any
underlying model like e.g. RPI. The RPI technique itself was used
independently e.g. in Refs.~\cite{Dio89,GGP90,Har91}.
In the context of quantum measurement theory, there is a fruitful
co-exitence of (a specific version of) RPI and WE techniques, as pointed
out recently \cite{Dioetal95}.

This work was supported by the grant OTKA No. 1822/1991.
\bigskip

\end{document}